\documentclass[aps,twocolumn]{revtex4}
\usepackage{amsfonts}
\usepackage{amsmath}
\usepackage{amssymb}
\usepackage{graphicx}

\begin{document}

\preprint{}
\title{Entanglement of Collectively Interacting Harmonic Chains: An effective%
\\
Two-Dimensional System}
\author{R.G. Unanyan$^{1,2}$ , M. Fleischhauer$^{2}$ and D. Bru\ss $^{1}$}
\affiliation{$^{1}$Institut f\"{u}r Theoretische Physik III, Heinrich-Heine-Universit\"{a}%
t D\"{u}sseldorf, D-40225 D\"{u}sseldorf, Germany\\
$^{2}$Fachbereich Physik, Technische Universit\"{a}t Kaiserslautern, 67663,
Kaiserslautern, Germany}

\begin{abstract}
We study the ground-state entanglement of one-dimensional harmonic chains
that are coupled to each other by a collective interaction as realized e.g.
in an anisotropic ion crystal. Due to the collective type of coupling, where
each chain interacts with every other one in the same way, the total system
shows critical behavior in the direction orthogonal to the chains while the
isolated harmonic chains can be gapped and non-critical. We derive lower and
most importantly upper bounds for the entanglement, quantified by the von
Neumann entropy, between a compact block of oscillators and its environment.
For sufficiently large size of the subsystems the bounds coincide and show
that the area law for entanglement is violated by a logarithmic correction.
\end{abstract}

\maketitle



Presently there is a growing interest in the interrelation between
entanglement and ground-state properties of many-body lattice models. For a
number of spin systems \cite{GVidal-PRL-2003} a strict correspondence
between the absence of criticality, the presence of an energy gap, and an
area law for the entanglement was established. The latter states that the
entanglement of a compact sub-set of lattice sites with the rest of the
system, measured by the von Neumann entropy, scales with the surface area of
the sub-set. For critical spin systems it was shown that an additional
logarithmic correction to the area law emerges. A similar relation between
criticality and entanglement was suggested for harmonic lattice models \cite%
{Bombelli-PRD-1986,Srednicki-PRL-1993}. In \cite%
{Plenio-PRL-2005,Plenio-PRA-2006} an area law was established for harmonic
lattice models in arbitrary dimensions with nearest-neighbor coupling which
have a gaped spectrum. For finite-range couplings in one dimension a
one-to-one correspondence between the validity of the area law and
non-criticality was established in \cite{RUFM}, and logarithmic corrections
were derived for critical systems.

Although the relation between criticality and entropy-area law seems rather
universal, there are a number of examples where this relation does not hold
\cite{Duer-PRL-2005,Plenio-PRA-2006}. Until now there is no general
understanding of the conditions for the validity of an entropy area law in
particular in higher dimensions \cite%
{GVidal-PRL-2003,Plenio-PRL-2005,Plenio-PRA-2006,Eisert-condmat-2006}. In
the present paper we discuss a specific gapless oscillator model with
dimension larger than one, for which an exact asymptotic expression for the
entropy can be obtained. Due to the collective nature of the interactions in
one spatial direction the system is critical and thus a violation of the
area law is expected. We here derive a lower and, most importantly, a tight
upper bound for the entropy and in this way obtain an exact form of the
correction term to the area law.

\begin{figure}[tbh]
\begin{center}
\includegraphics[width=8cm]{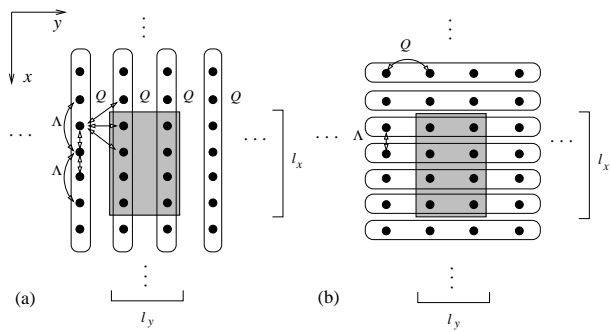}
\end{center}
\caption{(a) Collectively interacting strings of harmonic oscillators with
finite-range intra-chain coupling $\Lambda $ and collective inter-chain
coupling $Q$. The grey area indicates the sub-system I of oscillators. (b)
Alternative view: interacting strings with collective intra-chain coupling $Q
$ and finite-range inter-chain coupling $\Lambda $. }
\label{fig1}
\end{figure}

Let us consider a set of parallel harmonic chains (see Fig.\ref{fig1}) each
containing $n_{x}$ oscillators, with $n_{x}\rightarrow \infty $ in the
thermodynamic limit. We will refer to the direction parallel to the chains
as $x$-axis, and to the orthogonal direction as $y$-axis. The number of
parallel chains is denoted as $n_{y}$, again with $n_{y}\rightarrow \infty $
in the thermodynamic limit. The oscillators are described by the canonical
variables $\left( q_{i},p_{i}\right) $, where $i=1,2,...N$ ($N=n_{x}n_{y}$)
is a collective index that labels the oscillator. We adopt the following
notation: $i=1,...,n_{x}$ correspond to the oscillators in the first chain
with growing $x$ coordinate, $i=n_{x}+1,....,2n_{x}$ corresponds to
oscillators in the second chain and so on. We consider a quadratic
Hamiltonian of the form
\begin{equation}
H=\frac{1}{2}{\displaystyle\sum\limits_{i=1}^{N}}p_{i}^{2}+\frac{1}{2}{%
\displaystyle\sum\limits_{i,j=1}^{N}}V_{ij}q_{i}q_{j},  \label{Hamiltonian}
\end{equation}
with a coupling matrix $V$. We are interested only in a translationaly
invariant coupling, i.e. we assume that the matrix elements of $V$ depend
only on the difference of the $x$ coordinates and the difference of the $y$
coordinates. Hence $V$ is a block Toeplitz matrix. For oscillator systems
with a quadratic coupling of the form of eq.(\ref{Hamiltonian}) the ground
state
\begin{equation}
\Psi _{0}\left( \mathbf{q}\right) \sim \exp \left( -\frac{1}{2}\left\langle
\mathbf{q}\right\vert V^{1/2}\left\vert \mathbf{q}\right\rangle \right)
\label{ground-state}
\end{equation}
and all its properties, as e.g. the correlation length in position or
momentum space, are determined by the square root of $V$, where $\mathbf{q}%
=(q_{1},q_{2},\dots ,q_{N})$ is the vector of position variables. The ground
state can easily be determined if $V$ is the square of another matrix, which
we assume to be again a Toeplitz matrix,
\begin{equation}
V=Z^{2}/n_{y}.  \label{multiInteraction}
\end{equation}%
The factor $1/n_{y}$ is choosen such that the matrix elements of $V$ remain
finite in the limit $N\to \infty$. Assuming $Z$ to be a Toeplitz matrix
guarantees that the coupling $V$ is a Toeplitz matrix as well. We
furthermore consider $Z$ to be of the block-matrix form%
\begin{equation}
Z=\left[
\begin{array}{cccccc}
\Lambda & Q & Q & . & . & Q \\
Q & \Lambda & Q & . & . & Q \\
Q & Q & \Lambda & . &  &  \\
& . &  &  & . & Q \\
&  & . &  &  & Q \\
Q &  &  & Q & Q & \Lambda%
\end{array}%
\right] .  \label{Z matrix}
\end{equation}%
The elements of $Z$ are $n_x\times n_x$ matrices and characterize according
to eq.(\ref{ground-state}) correlations. The diagonal elements of $Z$
describe correlations within one chain, i.e. in $x$ direction, the
off-diagonal elements describe correlations between the chains. $\Lambda $
and $Q$ are both assumed to be Toeplitz matrices of finite range, i.e. their
matrix elements $\Lambda _{k}$ and $Q_{k}$, where $\Lambda _{k}\equiv\Lambda
_{k=|i-j|}=\langle i|\Lambda |j\rangle , $ vanish exactly for $k\geq R$. The
finite range of $\Lambda $ and $Q$ ensures that the interaction $V$ is of
finite range within the chains, while the form of $Z$ implies that $V$ is
constant orthogonal to the chains.
We assume furthermore that $\Lambda $, $Q$ and $\Lambda -Q$ are positive
definite matrices. A simple calculation shows that the ground state of $V$
is degenerate and in the thermodynamic limit $n_{x},n_{y}\rightarrow \infty $
has only one non-zero eigenvalue. This means that the total Hamiltonian, Eq.
(\ref{Hamiltonian}), is gapless. It should be noted however that the
collective nature of the interactions is not sufficient for a gapless
spectrum of the Hamiltonian.

Since all off-diagonal elements of $Z$ are identical, correlations between
oscillators do not depend on their distance in $y$ direction and the total
system is critical irrespective of the correlation properties within the
chains. Thus one expects that the entropy area law is broken. In fact one
can easily find a lower bound to the entropy by the following simple
argument: Let us consider a partition of the set of $N$ oscillators into a
compact sub-system I with $N_{0}=l_{x}l_{y}$ and a sub-system II with $%
N-N_{0}$ oscillators (see Fig.\ref{fig1}). If we now consider harmonic
chains in $y-$ rather than in $x-$ direction, the ``$y$''-chains couple to
each other with finite-range interaction $\Lambda $ (see Fig.\ref{fig1} b).
We thus have reason to assume that $S\ge l_{x}S_{0}$, where $S_{0}$ is the
entropy of a single ``$y$''-chain. Since the coupling within the chain is
now collective ($Q$), the ``$y$''-chain itself is critical and its entropy
scales as $S_{0}\sim \ln l_{y}$. Thus $S\ge l_{x}\ln l_{y}$ which in the
thermodynamic limit $\{l_x,l_y\}\to\infty$ is much larger than surface area $%
2(l_x+l_y)$. While it is easy to see that the area law is broken, it is
non-trivial to find an upper bound to the entropy and the exact form of the
correction term. This will be done in the following.

Using the spectral representation of $V$, the correlation matrices $V^{1/2}$
and $V^{-1/2}$ can be decomposed as
\begin{equation}
V^{1/2}=\left[ \left( \Lambda -Q\right) \otimes \mathbf{1}_{y}+n_{y}Q\otimes
\mathcal{P}_{n_y,n_y}\right] /\sqrt{n_y},  \label{correlationSpace}
\end{equation}%
and
\begin{eqnarray}
&&\quad V^{-1/2} = \Bigl\{( \Lambda -Q) ^{-1}\otimes \mathbf{1}_{y}+ \\
&& +\bigl[ ( \Lambda -Q+n_{y}Q) ^{-1}-( \Lambda -Q) ^{-1} \bigr] \otimes
\mathcal{P}_{n_y,n_y}\Bigr\}\sqrt{n_y},  \notag
\end{eqnarray}
where $\mathbf{1}_{y}$ is the unity matrix of size $n_{y}\times n_{y}$ and $%
\mathcal{P}_{nm}=|P_{nm}\rangle\langle P_{nm}|$ is the projector onto the
(in general non-normalized) vector
\begin{equation*}
\left\vert P_{nm}\right\rangle=\frac{1}{\sqrt{n}}\underset{m}{\Bigl(%
\underbrace{1,1,...1}}\Bigr) ^{T}.
\end{equation*}


Following Refs. \cite%
{Bombelli-PRD-1986,Srednicki-PRL-1993,Plenio-PRL-2005,Reznik}, the
von-Neumann entropy or the entropy of entanglement of the two compact parts
I and II can be calculated from a decomposition of $V^{1/2}$ into the two
subsystems. To this end we express $V^{1/2}$ and $V^{-1/2}$ in a block form
according to the two sub-systems by proper reordering of rows and columns
\begin{equation}
V^{-1/2}=\left[
\begin{array}{cc}
A & B \\
B^{T} & C%
\end{array}%
\right] ,\qquad V^{1/2}=\left[
\begin{array}{cc}
D & E \\
E^{T} & F%
\end{array}%
\right] .  \label{BlockForm}
\end{equation}%
Here $A$ and $D$ are $N_{0}\times N_{0}$ matrices describing correlations
within sub-system I, $C$ and $F$ are $\left( N-N_{0}\right) \times \left(
N-N_{0}\right) $ matrices describing correlations within sub-system II, and
the matrices $B$ and $E$ describe the correlations between them. The entropy
of entanglement is then given by the eigenvalues $\mu _{i}\geq 1$ of the
matrix product $A\cdot D$ \cite{Plenio-PRL-2005}:
\begin{eqnarray}
S &=&{\displaystyle\sum\limits_{i=1}^{N_{0}}f}\left( \sqrt{\mu _{i}}\right) ,
\label{Entropy} \\
{f}\left( x\right) &=&\frac{x+1}{2}\ln \frac{x+1}{2}-\frac{x-1}{2}\ln \frac{%
x-1}{2}.
\end{eqnarray}%
Despite the simplicity of its form, expression (\ref{Entropy}) cannot be
explicitly evaluated in general. Due to the special interaction matrix the
eigenvalues can however be evaluated in the thermodynamic limit:

From the spectral decomposition of $V^{1/2}$, eq.(\ref{correlationSpace}),
one easily finds that the subsystem matrices read
\begin{align}
A& =\left[ A_{0}\otimes \mathbf{1}_{l_{y}}+(A_{1}-A_{0})\otimes \mathcal{P}%
_{n_{y},l_{y}}\right] \sqrt{n_{y}},  \label{partition} \\
D& =\left[ D_{0}\otimes \mathbf{1}_{l_{y}}+n_{y}D_{1}\otimes \mathcal{P}%
_{n_{y},l_{y}}\right] /\sqrt{n_{y}},  \notag
\end{align}%
where $A_{0},A_{1}$ and $D_{0},D_{1}$ are $l_{x}\times l_{x}$ principal
submatrices of $(\Lambda -Q)^{-1},\left( \Lambda -Q+n_{y}Q\right) ^{-1}$,
and $(\Lambda -Q),Q$ respectively. For large $n_{y}$ one has
\begin{equation}
A\cdot D\approx \left( A_{0}\cdot D_{0}\right) \otimes \mathbf{1}%
_{l_{y}}+n_{y}\left( A_{0}\cdot D_{1}\right) \otimes \mathcal{P}%
_{n_{y},l_{y}}.  \label{AD}
\end{equation}%
Here we have used that $\mathcal{P}_{n_{y},l_{y}}^{2}=l_{y}/n_{y}\mathcal{P}%
_{n_{y},l_{y}}$ which scales as $1/n_{y}$ for fixed $l_{y}$ and is thus
negligible in the thermodynamic limit. Furthermore $\mathcal{P}_{n_{y},l_{y}}
$ has one nonzero eigenvalue ${l_{y}}/{n_{y}}$, which vanishes in the
thermodynamic limit ($l_{y}$ fixed and $n_{y}\rightarrow \infty $), and $%
\left( l_{y}-1\right) $ zero eigenvalues. Thus the $l_{x}l_{y}$ eigenvalues
of $A\cdot D$ can be decomposed into two sets. The first set consists of the
$l_{x}$ eigenvalues of $A_{0}\cdot D_{0}$ each of which occurs $(l_{y}-1)$
times:
\begin{eqnarray}
\mu _{1},\cdots ,\mu _{l_{y}-1} &=&\alpha _{1}\left( A_{0}\cdot D_{0}\right)
,  \notag \\
\mu _{l_{y}},\cdots ,\mu _{2(l_{y}-1)} &=&\alpha _{2}\left( A_{0}\cdot
D_{0}\right) ,  \label{first} \\
&\vdots &  \notag \\
\mu _{(l_{x}-1)(l_{y}-1)+1},\cdots ,\mu _{l_{x}(l_{y}-1)} &=&\alpha
_{l_{x}}\left( A_{0}\cdot D_{0}\right) .  \notag
\end{eqnarray}%
Here and in the following $\alpha _{k}\left( X\right) $ denotes the $k$th
eigenvalues of the matrix $X$. The total number of these eigenvalues is $%
l_{x}(l_{y}-1)$. The second set consists of the $l_{x}$ eigenvalues of $%
(A_{0}\cdot D_{0}+l_{y}\left( A_{0}\cdot D_{1}\right) )$
\begin{eqnarray}
\mu _{k} &=&\alpha _{k}\left( A_{0}\cdot D_{0}+l_{y}\left( A_{0}\cdot
D_{1}\right) \right) ,  \label{uppereigenvalues} \\
&&\qquad \text{for}\quad k=l_{x}(l_{y}-1)+1,...,l_{x}l_{y}.  \notag
\end{eqnarray}%
Expression (\ref{uppereigenvalues}) for the second set of eigenvalues can be
simplified using Lidskii's theorem \cite{Lidskii} which states: Let $X$ and $%
Y$ \ be $M$ -dimensional Hermitian matrices. Moreover let $\alpha _{k}\left(
X\right) ,\alpha _{k}\left( Y\right) $ and $\alpha _{k}\left( X-Y\right) $ ,
$k=1,...,M$ be the eigenvalues of $X,Y$ and $X-Y$ respectively in ascending
order $\{\alpha _{1}\left( X\right) \leq \alpha _{2}\left( X\right) \leq
...\leq \alpha _{M}\left( X\right)\} $. Then there exist numbers $w_{kj}\geq
0$, ($k,j=1,...,M$), such that $\sum_{k}w_{kj}=\sum_{j}w_{kj}=1$ and
\begin{equation}
\alpha _{k}\left( X\right) =\alpha _{k}\left( Y\right)
+\sum_{j=1}^{M}w_{kj}\alpha _{j}\left( X-Y\right) .  \label{Lidskii}
\end{equation}%
Equation (\ref{Lidskii}) implies that for sufficiently large $l_{y}$ the
eigenvalues of the matrix $A_{0}\cdot D_{0}+l_{y}\left( A_{0}\cdot
D_{1}\right) $ are
\begin{equation}
\alpha _{k}\left( A_{0}\cdot D_{0}+l_{y}\left( A_{0}\cdot D_{1}\right)
\right) \approx l_{y}\sum_{j=1}^{l_{x}}w_{kj}\alpha _{j}\left( A_{0}\cdot
D_{1}\right) .  \label{largelL}
\end{equation}%
%
%
%

An \textit{\ upper} bound to the entropy can be found by evaluating the sum
over the eigenvalues (\ref{first}) and (\ref{uppereigenvalues}) in eq.(\ref%
{Entropy}) separately
\begin{eqnarray}
S &=&S_{1}+S_{2} \\
&=&\sum_{j=1}^{l_{x}(l_{y}-1)}F\left( \sqrt{\mu _{j}}\right)
+\sum_{j=l_{x}(l_{y}-1)+1}^{l_{x}l_{y}}F\left( \sqrt{\mu _{j}}\right) .
\notag
\end{eqnarray}
Taking into account eq.(\ref{first}) one recognizes that $S_{1}$ is apart
from a prefactor $(l_{y}-1)$ formally equivalent to the von-Neumann entropy
of a linear oscillator chain of length $l_{x}$ with interaction $\tilde{V}%
=(\Lambda -Q)^{2}$
\begin{equation}
S_{1}=(l_{y}-1)\sum_{k=1}^{l_{x}}F\left( \sqrt{\alpha _{k}(A_{0}\cdot D_{0})}%
\right) .
\end{equation}
Since $\Lambda -Q$ was assumed to be strictly positive, the interaction $%
\tilde{V}$ has only nonzero eigenvalues and thus corresponds to a gaped
oscillator chain. As shown in \cite{Plenio-PRL-2005},\cite{RUFM} the entropy
of such a linear chain saturates in the thermodynamic limit, i.e it becomes
independent on the length $l_{x}$ of the chain. Thus we have in the
thermodynamic limit
\begin{equation}
S_{1}\leq l_{y}c_{1}.
\end{equation}
To obtain an upper bound to $S_{2}$ we use the inequality ${F}\left(
x\right) <1-\ln 2+\ln x$. This yields with eq.(\ref{largelL})
\begin{equation}
S_{2}<l_{x}(1-\ln 2)+\frac{1}{2}\sum_{k=1}^{l_{x}}\ln \left(
l_{y}\sum_{j=1}^{l_{x}}w_{kj}\alpha _{j}(A_{0}\cdot D_{1})\right) .
\end{equation}
To further evaluate the last term we make use of the convexity of the
logarithm together with the arithmetic mean inequality
\begin{eqnarray}
&&\frac{1}{2}\sum_{k=1}^{l_{x}}\ln \left(
l_{y}\sum_{j=1}^{l_{x}}w_{kj}\alpha _{j}(A_{0}\cdot D_{1})\right)  \notag \\
&&\quad \leq \frac{l_{x}}{2}\ln \left( \frac{l_{y}}{l_{x}}%
\sum_{j=1}^{l_{x}}\sum_{k=1}^{l_{x}}w_{kj}\alpha _{j}(A_{0}\cdot
D_{1})\right) \\
&&\quad =\frac{l_{x}}{2}\ln \left( \frac{l_{y}}{l_{x}}\sum_{j=1}^{l_{x}}%
\alpha _{j}(A_{0}\cdot D_{1})\right) ,  \notag
\end{eqnarray}
where we have used $\sum_{k}w_{kj}=1$ in the last step.

We now have to evaluate the remaining logarithm. For this we make use of the
fact that $\Lambda $ and $Q$ are regular (i.e. strictly positive) Toeplitz
matrices. Because of this, their elements can be obtained from the
non-negative spectral functions $\lambda \left( \theta \right) $ and $%
q\left( \theta \right) $ \cite{Szegoe} %
%
$\Lambda _{k} =\frac{1}{2\pi }\int\limits_{0}^{2\pi }\lambda \left( \theta
\right) \exp \left[ -ik\theta \right] d\theta$ , 

$Q_{k} =\frac{1}{2\pi }\int\limits_{0}^{2\pi }q\left( \theta \right) \exp %
\left[ -ik\theta \right] d\theta . $ 
%
%
Since we have assumed above that also $\Lambda -Q$ is strictly positive, the
functions $\lambda \left( \theta \right) $ , $q\left( \theta \right) $ are
strictly positive \ and $\lambda \left( \theta \right) >q\left( \theta
\right) $. In addition, we require also that $\left( \lambda \left( \theta
\right) -q\left( \theta \right) \right) ^{\pm 1}$ and $q\left( \theta
\right) $ have bounded derivatives of second order. As a consequence one
finds (see \cite{Szegoe} page 221)
\begin{equation}
\frac{1}{l_{x}}\left( \sum\limits_{j=1}^{l_{x}}\alpha _{j}\left( A_{0}\cdot
D_{1}\right) \right) \approx \frac{1}{2\pi }\int\limits_{0}^{2\pi }\frac{%
q\left( \theta \right) }{\lambda \left( \theta \right) -q\left( \theta
\right) }d\theta  \label{producteoplitz}
\end{equation}
which is a constant independent on $l_{x}$. Thus the desired upper bound to
the entropy for sufficiently large $l_{x},l_{y}$ is:%
\begin{equation}
S\leq c_{1}l_{y}+c_{2}l_{x}+\frac{l_{x}}{2}\ln l_{y}  \label{finalupperbound}
\end{equation}
where $c_{1},c_{2}$ are some constants independent of the size of the
subsystem.


A \textit{\ lower} bound to the entropy can be found from the inequality $%
F\left( x\right) \geq \ln x.$ This yields, with eq.(\ref{largelL}),
\begin{eqnarray}
S &\geq &\frac{(l_{y}-1)}{2}{\sum\limits_{k=1}^{l_{x}}\ln }\left[ \alpha
_{k}\left( A_{0}\cdot D_{0}\right) \right]  \label{lower1} \\
&&+\frac{l_{x}}{2}{\ln }(l_{y})+\frac{1}{2}\sum\limits_{k=1}^{l_{x}}{\ln }%
\left( \sum_{j=1}^{l_{x}}w_{kj}\alpha _{j}\left( A_{0}\cdot D_{1}\right)
\right) .  \notag
\end{eqnarray}
Making use of Jensen's inequality for concave functions $\ln \left(
\sum_{j}t_{j}\alpha _{j}\right) \geq \sum_{j}t_{j}\ln (\alpha _{j})$ and $%
\sum_{k}w_{kj}=1$ we find
\begin{eqnarray}
S &\geq &\frac{(l_{y}-1)}{2}{\sum\limits_{k=1}^{l_{x}}\ln }\left( \alpha
_{k}\left( A_{0}\cdot D_{0}\right) \right)  \label{lower2} \\
&&+\frac{l_{x}}{2}{\ln }(l_{y})+\frac{1}{2}\sum_{j=1}^{l_{x}}{\ln }\left(
\alpha _{j}\left( A_{0}\cdot D_{1}\right) \right) .  \notag
\end{eqnarray}

To evaluate the sums over the logarithms we employ Szeg\"{o}'s theorem \cite%
{Szegoe} for determinants of a Toeplitz matrices $T$. The theorem states:
for sufficiently large $l_{x}$%
\begin{equation*}
\ln \left( \det \left( T\right) \right) \approx
q_{0}l_{x}+\sum\limits_{k=1}^{\infty }k\left\vert q_{k}\right\vert ^{2},
\end{equation*}%
for regular spectral function $q\left( \theta \right) $. Here $q_{k}$ is
Fourier coefficients of $\ln q\left( \theta \right) $. Since moreover
\begin{eqnarray}
\sum_{j}\ln \left( \alpha _{j}(A_{0}\cdot D_{1})\right) &=&\ln \left(
\prod_{j}\alpha _{j}(A_{0}\cdot D_{1})\right)  \notag \\
&=&\ln \bigl[ \det (A_{0})\det (D_{1})\bigr] ,
\end{eqnarray}%
we eventually find the lower bound%
\begin{equation}
S\geq a_{1}l_{x}+a_{2}l_{y}+\frac{l_{x}}{2}{\ln }(l_{y}).  \label{lower}
\end{equation}%
Here $a_{1},a_{2}$ are constants independent of the size of the subsystem
and we have ignored an unimportant constant term.


By combining the two estimates (\ref{finalupperbound}) and (\ref{lower}) one
finds
\begin{equation*}
{c}_1\, l_x+c_{2}\, l_{y}+\frac{l_{x}}{2}\ln (l_{y})\, \geq\, S\, \geq \,
a_{1}\, l_x +a_{2} \, l_y+\frac{l_{x}}{2}{\ln }(l_{y}).  \label{Estimation}
\end{equation*}
Since both sides of this inequality have the same functional form, $S$
approaches for large $l_x, l_y$ the asymptotic value
\begin{equation}
S\approx \frac{l_{x}}{2}{\ln }(l_{y}),\qquad l_x, l_y \gg 1.
\label{Logcorrections}
\end{equation}
This is the main result of our paper. It shows that the entropy area law is
violated for a set of harmonic chains, which for themselves have a gaped
spectrum and are non-critical but become gapless by a collective interaction
between the chains. Both upper and lower bound to the entropy attain the
same logarithmic correction term to the area law.

A physical system that can be approximated by the model studied here is an
anisotropic ion crystal. In such a system the Coulomb-interaction in the
direction of the small lattice constant can in first approximation be
considered as collective, while the one in an orthogonal direction is of
finite range.

In conclusion, we derived an exact asymptotic expression for the
entanglement entropy of a critical system of interacting oscillators in more
than one dimension. We found that similar to one-dimensional systems \cite%
{RUFM} the entanglement area law is violated by a logarithmic correction
proportional to the surface area in the critical direction. To our knowledge
the system of collectively interacting harmonic strings considered here,
which is approximately realized e.g. in an anisotropic ion crystal,  is the
first nontrivial example of a critical two-dimensional system for which the
correction to the area law can explicitly be calculated.

This work was supported by the European Commission through the Integrated
Project FET/QIPC "SCALA".



\begin{thebibliography}{99}
\bibitem{GVidal-PRL-2003} G. Vidal, J. I. Lattore, E. Rico, and A. Kitaev,
Phys. Rev. Lett. \textbf{90}, 227902 (2003); V.E. Korepin, Phys. Rev. Lett.
\textbf{92, }096402 (2003); B.-Q. Jin and V.E. Korepin, J. Stat. Phys.
\textbf{116}, 79 (2004); A. R. Its, B. Q. Jin, and V. E. Korepin, J. Phys. A
\textbf{38}, 2975 (2005); P. Calabrese and J. Cardy, J. Stat. Mech. (2004)
P06002; L.-A. Wu, M.S. Sarandy and D.A. Lidar, Phys. Rev. Lett. \textbf{93},
250404 (2004); J. P. Keating and F. Mezzadri, Commun. Math. Phys.\textbf{\
252}, 543 (2004); P. Calabrese and J. Cardy, Phys. Rev. Lett. \textbf{94},
050501 (2005); A. Hamma, R. Ionicioiu, and P. Zanardi, Phys. Rev. A \textbf{%
71}, 022315 (2005)

\bibitem{Bombelli-PRD-1986} L. Bombelli, R. K. Koul, J. Lee, and R. D.
Sorkin, Phys. Rev. D \textbf{34}, 373 \ (1986).

\bibitem{Srednicki-PRL-1993} M. Srednicki, Phys. Rev. Lett. \textbf{71}, 666
(1993).

\bibitem{Plenio-PRL-2005} M. B. Plenio, J. Eisert, J. Drei\ss ig, and M.
Cramer, Phys. Rev. Lett. \textbf{94}, 060503 (2005).

\bibitem{Plenio-PRA-2006} M. Cramer, J. Eisert, M. B. Plenio, and J. Drei\ss %
ig, Phys. Rev. A \textbf{73}, 012309 (2006).

\bibitem{RUFM} R.G. Unanyan and M. Fleischhauer, Phys. Rev. Lett, \textbf{95}%
, 260604 (2005).

\bibitem{Duer-PRL-2005} W. D\"ur, L. Hartmann, M. Hein, M. Lewenstein, and
H.-J. Briegel, Phys. Rev. Lett. \textbf{94}, 097203 (2005).

\bibitem{Eisert-condmat-2006} M. Cramer, J. Eisert, and M.B. Plenio,
quant-ph/0611264

\bibitem{Reznik} A. Botero and B. Reznik, Phys. Rev. A.\textbf{70}, 052329

(2004).

\bibitem{Lidskii} B. V. Lidskii, Dokl. Akad. Nauk SSSR \textbf{74},
769(1950), see also: T. Kato, \textit{Perturbation theory for linear
operators}, Berlin, Spinger 1966.

\bibitem{Szegoe} U. Grenander and G. Szeg\"{o}, \textit{Toeplitz Forms and
their Applications} (University of California, Berkeley, 1958).

\
\end{thebibliography}
\end{document}